\documentclass[twocolumn,floats,amsmath,amssymb,prb,showkeys,a4paper]{revtex4}
\usepackage[english]{babel}
\usepackage[dvips]{graphicx}
\usepackage{bm}

\begin{document}
\title{\bf The nature of iron-oxygen vacancy defect centers in PbTiO$_3$}

\author{H. Me\v stri\'c} 
\author{R.-A. Eichel}
\email[corresponding author, fax: +49-6151-164347, e-mail: ]{eichel@chemie.tu-darmstadt.de}
\author{T. Kloss}
\author{K.-P. Dinse}
\author{So. Laubach}
\author{St. Laubach}
\author{P.C. Schmidt}

\affiliation{Eduard-Zintl-Institute, Darmstadt University of Technology, D-64287 Darmstadt, Germany}

\author{K.A. Sch\"onau} 
\author{M. Knapp}  
\author{H. Ehrenberg}
\affiliation{Materials Science, Darmstadt University of Technology, D-64287 Darmstadt, Germany}
\date{\today}

\begin{abstract}

The Fe$^{3+}$ center in ferroelectric PbTiO$_3$ together with an oxygen vacancy forms a
charged defect associate, oriented along the crystallographic
$c$-axis. Its microscopic structure has been analyzed in detail
comparing results from a semi-empirical {\it Newman superposition
model} analysis based on finestructure data and from calculations using density functional theory.\par

Both methods give evidence for a substitution of Fe$^{3+}$ for
Ti$^{4+}$ as an acceptor center. The position of the iron ion in
the ferroelectric phase is found to be similar to the B-site in the paraelectric phase. Partial
charge compensation is locally provided by a directly coordinated
oxygen vacancy.\par

Using high-resolution synchrotron powder diffraction, it was
verified that lead titanate remains tetragonal down to 12 K,
exhibiting a $c/a$-ratio of 1.0721.\par
\end{abstract}

\date{22 February 2005}

\maketitle

\section{Introduction}
Piezoelectric lead titanate oxide (PbTiO$_3$, PT) is of considerable
scientific and technical interest because of its extraordinary
electromechanical properties leading to the development of novel
devices, such as nonvolatile memories, detectors, sensors or
actuators \cite{Lines,Swa90,Sco98}. A variety of ferroelectric
properties can be controlled by replacing the lead or titanium
cations with rare-earth or transition-metal ions. In general, the
effect of aliovalent dopants on PbTiO$_3$, such as Fe$^{3+}$,
leads to the creation of oxygen vacancies ($V_{\rm O}^{\bullet
\bullet}$) (The Kr\"oger and Vink notation is used to
designate the charge state of the defect with respect to the
neutral lattice.) for charge compensation, which are expected to
have a major impact on properties and performance of ferroelectric
compounds. However, it is still not certain at which coordination
sphere the charge compensation takes place, and  because the
oxygen vacancies are known to be the dominant charge carriers in
this class of compounds this is an issue of technological and
scientific importance \cite{Smyth}. Beside charge compensation
provided by $V_{\rm O}^{\bullet \bullet}$, intrinsic doubly
negatively charged $V''_{\rm Pb}$ centers have been proposed as
additional charge compensating mechanism \cite{War9697}, these
vacancies, however, being rather immobile and almost of no
importance for charge transport.\par

When dealing with dopants on a sub-percentage level, a sensitive
test of the local environment around the functional center can
only be provided by electron paramagnetic resonance (EPR). In case
of paramagnetic Fe$^{3+}$ dopants, the sextet spin ground state
is most influenced by the ligand-field originating from the
nearest-neighbor O$^{2-}$ ions. The resulting finestructure (FS)
interaction is a measure of the local symmetry. Various structure
models have been proposed such as Fe$_{\rm Ti}^{\prime}-V_{\rm
O}^{\bullet \bullet}$ defect associates, not-coordinated 'free'
Fe$_{\rm Ti}^{\prime}$ centers, as well as off-center shifted iron
centers. Previous EPR studies on crystalline samples reported FS
parameters spanning a large range \cite{Fe_EPR}. Structural
information about the dopant site in PbTiO$_3$ and related
compounds can be based on modelling the magnitude of the FS
parameter, and therefore its accurate determination is an important
issue. Because most samples are provided in polycrystalline form,
it was important to demonstrate that FS parameters can reliably be
determined even for polycrystalline compounds by invoking
high-frequency-EPR \cite{RAE04c}.\par

Structure modelling can be performed using different levels of
sophistication. For instance, the FS parameter can be analyzed in
terms of the semi-empirical {\it Newman superposition model} ({\it
NSM}) \cite{New71}. Within this model, values for local
distortion, i.e., atomic displacements from the ideal crystal
structure in the vicinity of the Fe$^{3+}$ center, are derived by
comparing calculated and observed FS parameters. A detailed study
was performed already 25 years ago by Siegel and M\"uller
\cite{Sie79a,Sie79b}, in which the distortion of the cubic
high-temperature unit cell in the ferroelectric phase was taken as
decisive parameter controlling the FS splitting at the iron site.
Over the past decades, the {\it NSM} parametrization has been
refined by applying the method to many examples of paramagnetic
centers in perovskite-type crystals. Examples include iron-doped
PbTiO$_3$, SrTiO$_3$ and BaTiO$_3$ \cite{Sie79a,Sie79b}, LiTaO$_3$
\cite{Yeo01}, as well as chromium-modified PbTiO$_3$ \cite{Erd03}
and manganese in BaTiO$_3$ \cite{Boe00}.\par

\begin{figure}[ht]
 \includegraphics[width=\columnwidth]{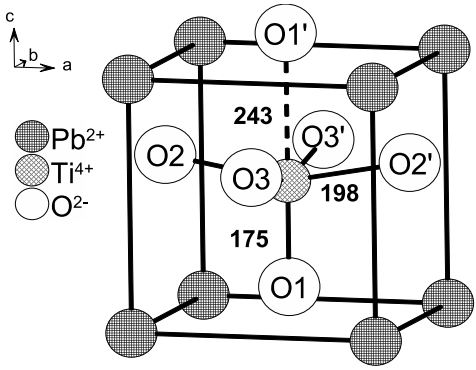}
 \begin{center}
  \caption{Schematic representation of the tetragonal unit cell of PbTiO$_3$ with designation of the different oxygen sites. The experimental atomic distances for the Ti-site to the surrounding oxygen octaeder are given in pm.}
  \label{3D_structure}
 \end{center}
\end{figure}

Second, an alternate approach offering additional microscopic
information, is provided by calculating the equilibrium structure
using density-functional theory ({\it DFT}). In recent years numerous
calculations have been performed on ferroelectric materials in
order to explain the microscopic mechanisms of spontaneous
polarization \cite{Kin9394,Cohen} and phase diagrams
\cite{Zho94,Wag97}, or to determine the full piezoelectric stress
tensor and dynamical charges \cite{Sag9899}. In particular, the
role of defects and defect dipoles has been highlighted. Oxygen
vacancies were shown to pin at 180$^{\circ}$-domain walls,
confirming the tendency of these defects to migrate to the walls
\cite{He03}. It was also derived that lead vacancies have stable
charges ranging from $-2$ to $-4$ and thus being an effective
acceptor in lead titanate \cite{Poe00}. The dipole moment $V_{\rm
O}^{\bullet \bullet}-V''_{\rm Pb}$ of  di-vacancies was
calculated, demonstrating that the increased value may be an
important source of local polarization and electric fields
\cite{Coc04}. Another defect dipole originates from interstitial
hydrogen impurities in lead titanate, which were found to bind to
oxygen and to act as shallow donor impurities. The H-O dipole
increases the polarization and the barrier for reversing the
defect dipole, which can give rise to imprint \cite{Par00}.\par

In this work, we exploit the recently accurately determined FS
parameter \cite{RAE04c} as basis for structure modelling by performing an {\it NSM}
analysis. Second, we compare the results obtained from this
semi-empirical approach with results from {\it DFT} calculations, thus
providing a sound basis for the postulated structural relaxation
at the impurity site. As a result, we present a refined structural
model of the iron functional center in lead titanate.

\section{Theoretical methods}
\subsection{Newman superposition model}
With its five unpaired electrons, the free Fe$^{3+}$ ion possesses
a half-filled $3d$ shell and can be described as orbital singlet.
Its ground state configuration is $^6S_{5/2}$ ($S = \frac{5}{2}$)
and the six-fold spin degeneracy can be lifted by the FS
interaction and an external magnetic field. Neglecting hyperfine
interaction, because the magnetic active isotope $^{57}$Fe is
present in only 2.2 \% natural abundance, an approximate
spin-Hamiltonian for this high-spin system can be written as

\begin{equation}
 \label{Hamiltonian_general}
 {\cal H} = \beta_e {\bf B_0 \cdot g \cdot S} + \sum_{k,q} B_k^{q}O_k^{q}(S_x,S_y,S_z)
\end{equation}

Here, $\beta_e$ denotes the Bohr magneton, ${\bf B_0}$ is the
external magnetic field, ${\bf g}$ the electron ${\bf
g}$-matrix, $B_k^{q}$ are the FS Hamiltonian coefficients and
$O_k^{q}$ are the extended Stevens spin operators \cite{Abr70,Rud99}. The first term represents the electronic
Zeeman interaction and the second term is the effective FS Hamiltonian, describing the interaction of the crystal field with
the paramagnetic ion. The rank $k$ in the Hamiltonian must be even, and $k$
is restricted by $k \leq 2S$ and $q \leq k$, allowing terms up to $k
= 4$ for $S=\frac{5}{2}$. However, parameters $b_k^q$ only up to
second rank ($k = 2$) will be used, because terms of fourth rank were shown to be at
least two orders of magnitude smaller than those of second rank
\cite{Fe_EPR}(c-e). Parameters $b_k^q$ are related to the standard
FS parameters $B_k^q$ by means of the scaling factor as $B_k^q =
f_k b_k^q$. For $k = 2$ one finds $f_2=\frac{1}{3}$. The
coefficients $b_k^q$ are related to the conventional spectroscopic FS parameters
$D$ and $E$ by $b_2^0 = 3B_2^0 = D$ and $b_2^2 = 3B_2^2 = 3E$.\par

The FS term in (\ref{Hamiltonian_general}) lifts the degeneracy of
eigen states even in the absence of an external magnetic field.
There is always a choice of spin-operator coordinate system with
respect to the crystal axes for which the ratio $b_2^2/b_2^0$ lies
in the interval $[0,1]$, analogous to the condition $E/D \le
1/3$.\par

In case the symmetry of the paramagnetic site is axial, the FS
coefficients in its eigen frame reduce to $b_2^0 \ne 0$,
$b_2^2 = 0$. Furthermore, the ${\bf g}$-matrix for $S$-state ions
usually has very small anisotropy and can be treated as isotropic.
The resulting scalar $g$ value is expected to be close to that of
the free electron $g_e$. For Fe$^{3+}$ we therefore set $g_{\rm
iso} = 2.002$ and then the spin-Hamiltonian used for numerical spectrum
simulation of the experimental data reduces to

\begin{equation}
 \label{SimpleHam}
 {\cal H} = \beta_e g_{\rm iso} \; {\bf B_0 \cdot S} + b_2^0 \left[ S_z^2 - \frac{1}{3} S(S+1) \right]
\end{equation}

The {\it Newman superposition model} \cite{New71} allows for
an analysis of positions of neighboring ions (ligands) around the
Fe$^{3+}$ center, given the single-ligand contributions
$\bar{b}_k(R_i)$ are known. The essential assumption of the {\it NSM} is
that the spin-Hamiltonian parameters for a paramagnetic ion can be
constructed by a superposition from individual contributions of
separate neighboring ligands. The contribution of next nearest
neighbor ions as well as interaction between the ligands are
ignored. The {\it NSM} expression for the zero-field splitting
parameters can be formulated as follows

\begin{equation}
 \label{SM}
 b_k^q = \sum_{i}\bar{b}_k(R_i) \; K_k^q(\theta_i,\phi_i)
\end{equation}

Here, $i$ denotes every ligand, $\bar{b}_k$ is the contribution
from each single ligand, $R_i$ is the distance between the $i$th
ligand and the paramagnetic ion, $\theta_i$ and $\phi_i$ are the
polar and axial angles between $R_i$ and the symmetry axis of the
paramagnetic center, and $K_k^q(\theta_i,\phi_i)$ are spherical harmonic functions of rank
$k$ of the polar angles, listed in \cite{New71}(f). For the axial
second-rank parameter one has

\begin{equation}
 \label{CF}
 K_2^0 = \frac12 (3\cos^2\theta - 1)
\end{equation}

For the short-range distance dependence of the single-ligand
contribution, the following empirical power law was established

\begin{equation}
 \label{power_law}
 \bar{b}_k(R_i) = \bar{b}_k(R_0)\left( \frac{R_0}{R_i} \right)^{t_k}
\end{equation}

The critical exponent parameter $t_k$ is specific to a particular
ion-ligand system. For a given ligand, the {\it intrinsic
parameters} $\bar{b}_k (R_i)$ are determined by the nature of the
ligand and the covalency of the bond, which is assumed to depend
exclusively on the bond lengths $R_i$. Because $\bar{b}_k(R_i)$ in
this model depends only on the ligand and its distance, and not on
other properties of the host crystal, this parameter can be
obtained from data of the same ion - ligand complex measured in
other host crystals.\par

If the molecular coordinate system coincides with the principal
axes system of the FS tensor, all off-diagonal elements vanish.
For the Fe$^{3+}$ ion at a position of tetragonal site symmetry ($4mm$), the {\it NSM} can be used in its truncated form \cite{Sie79a}

\begin{equation}
 b_2^0 = \bar{b}_2(R_0) \frac{3}{2} \sum_i \left( \frac{R_0}{R_i} \right)^{t_2} \left( \cos^2 \theta_i - \frac{1}{3} \right)
\end{equation}

The strategy followed is to calculate the FS parameter by using
the {\it NSM} for different conceivable structural arrangements,
and to compare the calculated values with the experimental data.
The experimental value for $b_2^0$ has been determined reliably
\cite{RAE04c}, and thus the interception points between the
calculated curve and the experimental value is used for a
prediction of actually realized local structure. Previously, the
extremum of the calculated $b_2^0$ dependence was used to obtain
an approximate ion position, although a correlation between the
ZFS extremum and the ion equilibrium position cannot be derived
from first principles. \cite{Sie79a,Sie79b}.

\subsection{Density functional theory ({\it DFT}) modelling}
The electronic structure of Fe:PbTiO$_3$ was calculated using two
different {\it DFT} methods. First, the local relaxation of the
structure around the defects, Fe$_{\rm Ti}$, and V$_{\rm O}$, is
studied by the Vienna Ab initio Simulation Package (VASP)
\cite{T1}, as applied recently to study Pb-O vacancies in
PbTiO$_3$ \cite{Coc04}. However, instead of using ultrasoft
pseudopotentials, Projector Augmented Waves (PAW) have been
applied \cite{T2,T3}. The identical basis set has been used,
namely Pb($5d$,$6s$,$6p$), Ti($3s$,$3p$,$3d$,$4s$,$4p$), and
O($2s$,$2p$), as well as the LDA exchange potential \cite{T4}, and
a $4 \times 4 \times 4$ k-mesh. Second, the electron density
distribution $\rho(r)$ has been computed by the NFP (new full
potential) program package \cite{T5,T6,T7}, which is a variant of
the LMTO (linear muffin-tin orbitals) \cite{T8} procedure using a
minimal basis set.

\begin{figure}[ht]
  \includegraphics[width=\linewidth]{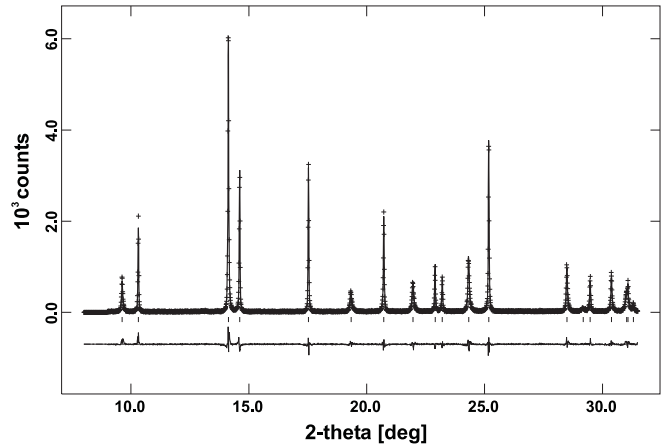}
 \begin{center}
  \caption{Observed and calculated diffraction profiles for PbTiO$_3$, obtained from high-resolution synchrotron powder diffraction at 12 K (top) together with their difference curve (bottom); $\lambda = 0.0699988$ nm.}
  \label{xray_PT}
 \end{center}
\end{figure}

\begin{table*}[ht]
 \label{X_ray_results}
 \renewcommand{\arraystretch}{2.0}
 \begin{center}
  \caption{Lattice parameters ($a,b = 389.0673(11)$ pm, $c = 416.7553(30)$ pm), cell volume ($0.0630856(5)$ nm$^3$) and relative atom coordinates (upper rows) for PbTiO$_3$ at 12 K, obtained from high-resolution synchrotron powder diffraction. $U_i$ is the mean square of atomic displacement. Also listed are the predicted values from {\it DFT} calculations with lattice parameters $a = b = 386.4$ pm, $c = 404.2$ pm (lower rows). Pb was kept fixed at the origin for both, Rietveld refinement and {\it DFT} calculations.}
 \end{center}
 \renewcommand{\arraystretch}{2.0}
  \begin{center}
   \begin{tabular}{|l|l|l|l|l|l|l|} \hline
    atom & & $X$ & $Y$ & $Z$ & $U_i \cdot 100$ & site symmetry \\ \hline \hline
    Pb      & exp.      & 0.000000 & 0.000000 & 0.000000   & 0.820(27) & 4MM(001) \\
            & {\it DFT} & 0.000000 & 0.000000 & 0.000000   &           &          \\ \hline
    Ti      & exp.      & 0.500000 & 0.500000 & 0.5415(14) & 0.66(10)  & 4MM(001) \\
            & {\it DFT} & 0.500000 & 0.500000 & 0.5346     &           &          \\ \hline
    O1      & exp.      & 0.500000 & 0.500000 & 0.1239(19) & 0.38(19)  & 4MM(001) \\
            & {\it DFT} & 0.500000 & 0.500000 & 0.0904     &           &  \\ \hline
    O2 / O3 & exp.      & 0.000000 & 0.500000 & 0.6288(13) & 0.38(19)  & MM2(001) \\
            & {\it DFT} & 0.000000 & 0.500000 & 0.6032     &           & \\ \hline
   \end{tabular}
  \end{center}
\end{table*}

\section{Results}
\subsection{Low-temperature X-ray data}
\label{X_ray} The low temperature crystal structure of lead
titanate has been the subject of some controversy over the past 50
years. Several X-ray and optical studies as well as dielectric
measurements of PT samples indicated small anomalies in the
temperature regions around 173 K and 123 K. These anomalies were
detected as changes in the negative volume expansion coefficient
in combination with superlattice reflections in powder samples
\cite{Kob55} below 173 K and as changes in lattice parameters and
birefringence in single crystals around 183 K. They were
interpreted as a possible second order phase transition to a
different  tetragonal, later corrected into an antiferroelectric
orthorhombic phase \cite{Kob83}. Moreover, this earlier claimed
low-temperature tetragonal phase was assumed to transform into a
further tetragonal phase around 113 K \cite{Ike69}, which does not
fit into the picture of having an orthorhombic phase present at
this temperature range. However, neither in neutron powder
profiles \cite{Gla78}, nor in X-ray dilatometric and optical
measurements using PT single crystals, additional evidence was
seen for a structural transition at low temperatures \cite{Mab79}.
First-principle studies \cite{Gar96} show that the
thermodynamically stable phase at low temperatures is the
tetragonal one, as all the unit-cell preserving distortions at low
temperature have positive elastic constants and as there are no
other mechanical instabilities present that could cause a
transition to a lower symmetry group.\par

To identify the low-temperature structure and to obtain reference
values for the interpretation of EPR data recorded at low
temperatures, a high-resolution synchrotron powder diffraction
experiment was carried out in reflection geometry at B2, Hasylab
in Hamburg, Germany (figure \ref{xray_PT}). The measurement was
performed at 12 K using a He closed-cycle cryostat, equilibrating
the temperature in a low-pressure helium atmosphere. The powdered
sample was glued onto a Si (711) low-background wafer and measured
using an incident beam of wavelength $0.0699988$ nm in combination
with an analyzing crystal and a scintillation counter
\cite{Kna04}.\par

For Rietveld refinement, the general structure analysis system
(GSAS) \cite{Lar94} was used. As the peak profiles of the
ferroelectric material are quite complex, the profiles were fitted
using the incorporated generalized model for anisotropic peak
broadening \cite{Ste99}. The structural model could be refined
with the symmetry of $P 4mm$, keeping the lead positions fixed at
the origin. No superlattice reflections were detected. The $c/a$ -
ratio is 1.0721 at this temperature. Lattice parameters and atom
coordinates obtained are given in table \ref{X_ray_results}.\par

The detected low-temperature $P 4mm$ symmetry is consistent with the
interpretation of the EPR spectra, for which a FS tensor of axial
symmetry was used, thus indicating that no orthorhombic phase is
present \cite{RAE04c}. Therefore it can be concluded that
PbTiO$_3$ remains tetragonal down to 12 K, as already assumed
earlier \cite{Gar96}.

\begin{figure*}[ht]
 \begin{center}
  \includegraphics[width=\columnwidth, angle=270]{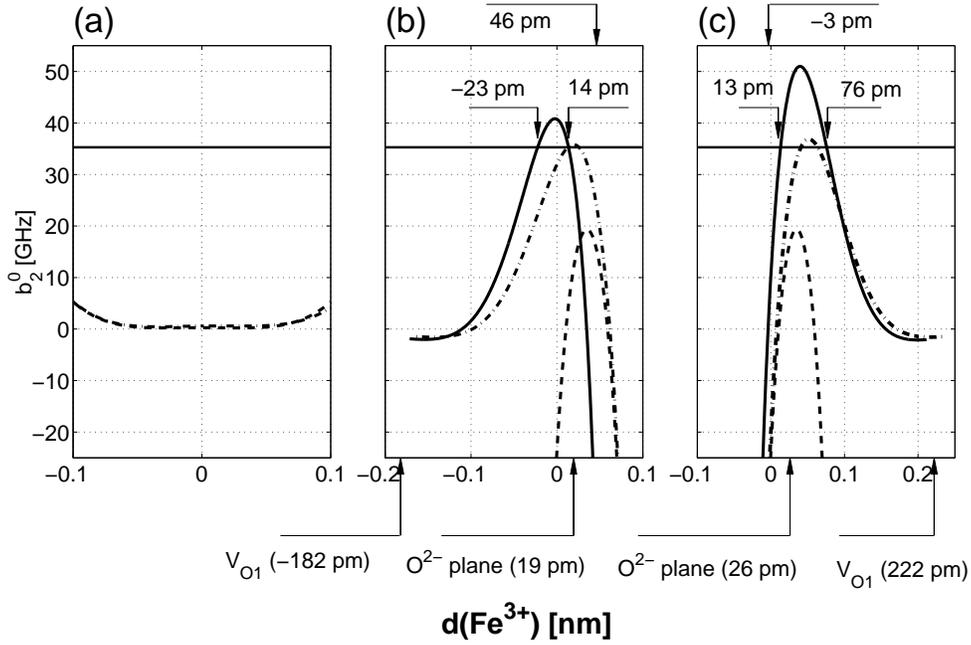}
  \caption{Comparison of the experimentally obtained axial FS parameter $b_2^0 = 35.28$ GHz (solid horizontal line) for the Fe$^{3+}$ center in PbTiO$_3$ \cite{RAE04c} with calculated values for $b_2^0$ obtained from the {\it NSM} analysis in dependance of the displacement $d$ from the Ti$^{4+}$ position. The positions of the oxygen vacancies, of the center of the oxygen octahedra, as well as of the DFT-calculated shifts of the Fe$^{3+}$ ion are indexed by arrows as reference. (a) - substitution of the Fe$^{3+}$ ion either at the A-site. (b,c) - substitution of the Fe$^{3+}$ ion at the B-site with oxygen vacancies either at O1 or O1'. The model for 'free iron', i.e. without an oxygen vacancy in the nearest neighbor shell, is represented in all models by a dashed line and the model for the iron-oxygen vacancy associate by a dash-dotted line in case of static oxygen position and by a solid line for the model using relaxed oxygen positions.}
  \label{NSM}
 \end{center}
\end{figure*}

\subsection{Superposition-model analysis}
\label{NSM_analysis} The intrinsic {\it NSM} parameters $t_2$ and
$\bar{b}_2 (R_0)$ have not yet been determined for iron centers in
lead titanate. However, they may be adopted from similar single
crystals having equivalent Fe$^{3+}-$O$^{2-}$ bonds in octahedral
coordination. For Fe$^{3+}$ in MgO at the central site with
octahedral oxygen coordination, the following set of {\it NSM}
parameters was reported: $\bar{b}_2 = -12.3514$ GHz, $t_2 = 8$,
$R_0 = 210.1$ pm \cite{Sie79a}. The ionic positions of the nearest
oxygens were taken from the X-ray data at 12 K (cf. section
\ref{X_ray}). Because the site symmetry defines the direction of
FS-tensor axes, the principal FS axis of the Fe$^{3+}$ center was
chosen along [0 0 1].\par

The common reference structure for undistorted cubic PbTiO$_3$ is
the perovskite structure, in which Ti$^{4+}$ ions are octahedrally
coordinated by oxygen ions and are positioned at the center of the
unit cell. The Pb$^{2+}$ ions are twelvefold coordinated by
oxygens and are located at the corners of the cube, whereas the
O$^{2-}$ ions are centered on each face of the unit cell. In the
ferroelectric tetragonal phase, the oxygen octahedron is elongated
along [0 0 1] and also shifted by 54 pm with respect to the
Pb$^{2+}$ position, the titanium ion being displaced from the
center of cell along [0 0 1] by 17 pm, respectively. As a result,
the titanium ion occupies an {\it off-center} position within the
shifted oxygen octahedron (cf. table \ref{X_ray_results}, figures
\ref{3D_structure}, \ref{structure}).\par

Iron is believed to substitute Ti$^{4+}$ as Fe$^{3+}$ at the
B-site in the ABO$_3$ perovskite structure. In order to verify
this hypothesis, in the calculation we also considered the
possibility that it substitutes at the A-site for Pb$^{2+}$.
Furthermore, to decide whether the oxygen vacancy is situated in
the first coordination sphere or if it is located more distantly,
two more structure models were investigated, with and without
oxygen vacancy in nearest neighbor positions. The model of a
directly coordinated oxygen vacancy can be realized in two
different arrangements, in which the vacancy may substitute either
for the apical O1 or O1' position (cf. figure \ref{3D_structure}).
A vacancy coordination at the equatorial O2 and O3 positions can
be excluded as the orientation of the Fe$'_{\rm Ti}-V_{\rm
O}^{\bullet \bullet}$ defect dipole has to be along the $c$-axis,
due to the EPR results that indicate axial symmetry at the iron
center \cite{RAE04c}. For orientations of the oxygen vacancy along
the $a$ and $b$ axes, the FS tensor would be of only orthorhombic
symmetry. For a final model, the effect of a likely relaxation of
oxygen ion positions in case of a directly coordinated oxygen
vacancy has been taken into account. The relaxed ionic positions
were obtained from the {\it DFT} calculations (cf. section
\ref{ab_initio}).\par

The position of the oxygen octaeder is defined using the crystal
coordinate system given in table \ref{X_ray_results}. Here, the
oxygen vacancy is located either at (0.5000 0.5000 0.1239), or at
(0.5000 0.5000 1.1239) in fractional coordinates, and the
reference position of iron is at the titanium position in the
undoped system at (0.5000 0.5000 0.5415). For the calculation of
$b_2^0$, the position of Fe$^{3+}$ was varied along [0 0 1], the
parameter $d$ defining the shift of the ion. A positive sign of
$d$ defines a shift towards the O1' oxygen, i.e. $d = z \times
416.8$ pm.\par

The results are presented in figure \ref{NSM}, depicting the
dependence of calculated axial FS parameters $b_2^0$ on $d$ for
the different structural models. In this figure, the
experimentally  obtained value for $b_2^0$ is represented by a
solid horizontal line. As reference, the positions of the
corresponding oxygen vacancies $V_{O1}^{\bullet \bullet}$,
$V_{O1'}^{\bullet \bullet}$, the position of the equatorial plane
of oxygens through O2, O3, as well as the positions for the iron
as obtained by the DFT calculations (cf. section \ref{ab_initio})
are indicated by arrows.\par

Four conclusions immediately emerge from the NSM calculations: as
compared to earlier studies for the iron center in lead titanate
\cite{Sie79a,Sie79b}, in which static oxygen positions were used,
the inclusion of relaxed positions for the equatorial oxygens
towards the vacancy yields substantially refined and quite
different curves for $b_2^0$. Furthermore, the extremum of the
$b_2^0$ dependence is no a-priori viable position for the iron
displacement as has earlier been exploited as first guess
\cite{Sie79a,Sie79b}; the intersection points with the
experimental value, not necessarily coinciding with the extremum,
should be used instead. Finally, agreement with the experimental
value \cite{RAE04c} can be obtained only with a model, in which
Fe$^{3+}$ is substituted at the B-site with a directly coordinated
oxygen vacancy, thus forming an Fe$'_{\rm Ti}-V_{\rm O}^{\bullet
\bullet}$ defect associate.\par

In the pure, unsubstituted ferroelectric phase, Ti$^{4+}$ is
displaced considerably from the center of the oxygen octahedron
(cf. table \ref{X_ray_results}). In contrast, there are two
predicted positions for the Fe$_{\rm Ti}^{3+}$ ion with an
adjacent oxygen vacancy for each of the two models: for the
Fe$'_{\rm Ti}-V_{\rm O1}^{\bullet \bullet}$ associate the iron ion
is either displaced by $14 \pm 5$ pm from the
Ti$^{4+}$ position away from the vacancy towards the center of the
truncated octahedron, or it is shifted considerably by $23 \pm 5$
pm towards the vacancy. For the alternative Fe$'_{\rm Ti}-V_{\rm
O1'}^{\bullet \bullet}$ associate, the Fe$_{\rm Ti}^{3+}$ ion both
times is shifted towards the vacancy, where one position is close
to the plane of equatorial oxygens at $13 \pm 5$ pm and the other
position is at $76 \pm 5$ pm. In order to distinguish between
these positions, further information as provided by the {\it DFT}
calculation is needed. The finally preferred structure (vide
infra) of the Fe$'_{\rm Ti}-V_{\rm O}^{\bullet \bullet}$ defect
associate from the {\it NSM} calculations is presented in figure
\ref{structure}.

\subsection{{\it DFT} calculations}
\label{ab_initio}
\subsubsection{{\it DFT} investigations of undoped PbTiO$_3$}
\begin{figure*}[ht]
 \begin{center}
  \includegraphics[width=1.5\columnwidth]{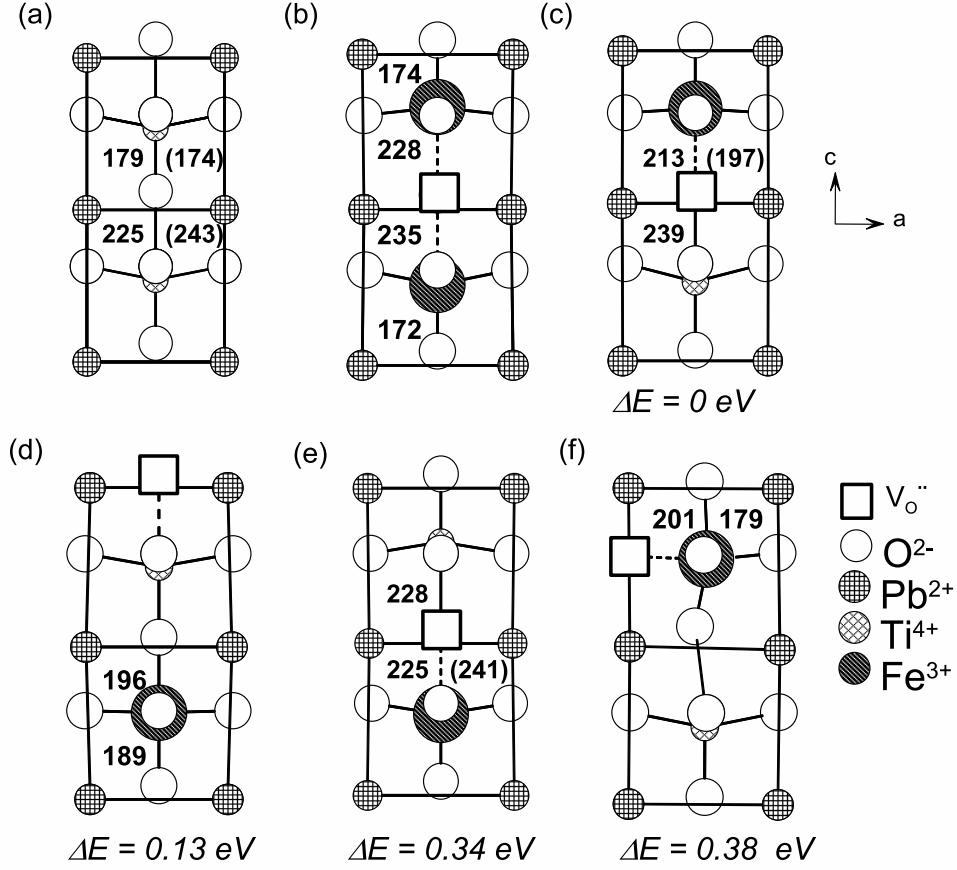}
  \caption{Comparison between the different structure models for Fe:PbTiO$_3$ deduced from the theoretical and experimental investigations. The given numbers indicate atomic distances (in pm) calculated by {\it DFT} (cf. section 3.3.2) and the bracketed values are experimentally and semi-emprically obtained values. The relative energies for comparable models are shown below the figures. \newline
(a) undoped reference structure obtained from DFT calculations and from synchrotron powder difraction (cf. table \ref{X_ray_results});\newline
(b) Fe$^\prime_{\rm Ti}$-V$_{\rm O}^{\bullet \bullet}$- Fe'$_{\rm Ti}$ complex in Pb$_8$Ti$_6$Fe$_2$O$_{23}$ oriented along $c$-axis (model A1);\newline
(c) Fe$^\prime_{\rm Ti}$-V$_{\rm O}^{\bullet \bullet}$ defect associate in [Pb$_8$Ti$_7$FeO$_{23}]^+$ oriented along $c$-axis where the O1 oxygen is missing (shorter bond, model C1). The bracketed values are obtained with Newman superposition model on the basis of measured FS data. This model is the energetically most favored among models with one oxygen vacancy.\newline
(d) Fe$^\prime_{\rm Ti}$ and V$_{\rm O}^{\bullet \bullet}$ as isolated defects in [Pb$_8$Ti$_7$FeO$_{23}]^+$ (model C3); \newline
(e) Fe$^\prime_{\rm Ti}$-V$_{\rm O}^{\bullet \bullet}$ defect associate in [Pb$_8$Ti$_7$FeO$_{23}]^+$ oriented along $c$-axis where the oxygen O1$^\prime$ is missing (longer bond, model C2); \newline
(f) Fe$^\prime_{\rm Ti}-V_{\rm O}^{\bullet \bullet}$ defect associate in [Pb$_8$Ti$_7$Fe$_2$O$_{23}]^+$ oriented along $a$-axis where the oxygen O2 is missing (model C4); \newline
Models B1-B4 which consider defects in charged Pb$_8$Ti$_7$Fe$_2$O$_{23}$ are in the same energetical sequence.}
  \label{structure}
 \end{center}
\end{figure*}

First, the iron-free host lattice of PbTiO$_3$ was studied. Using
the VASP code, lattice parameters $a = 386.6$ pm and $c = 403.9$
pm were obtained, which is in agreement with earlier calculations
\cite{Mey02}. These values are smaller than the experimental
values as expected from LDA calculations. The percentage
difference between the experimental and calculated values are
different for the $a$ and $c$ direction resulting in a calculated
$c/a$ strain of 1.04 compared to the experimental value of 1.07.
The calculated Ti-O bonds are given in figure
\ref{structure}(c).\par

Next, one oxygen vacancy was inserted in the $2 \times 2 \times 2$
supercell without providing charge compensation, i.e., the
electronic structure of Pb$_8$Ti$_8$O$_{23}$ was calculated for a
determination of the local relaxation of the ions around the
vacancy. As shown in figure \ref{structure} (d), one finds that
the nearest neighbor Ti ions move away from the vacancy site,
which was also predicted in a previous LDA calculation
\cite{Par98}. If the vacancy is located at the O1 site (cf. figure
\ref{3D_structure}), the Ti moves towards O1' and the Ti-O1'
distance decreases in the LDA simulation from $224.5$ pm (in the
perfect lattice) to $193.2$ pm. The Ti ion closest to the vacancy
in positive c-direction moves from its original position by $d = +
25.4$ pm while the nearest Ti ion on the opposite side of the
vacancy is displaced by $d = -3.9$ pm. If the vacancy is located
at the O2 site, Ti moves towards the O2' site and the Ti-O2' bond
distance decreases from $195.3$ pm to $187.6$ pm.

\subsubsection{{\it DFT} investigations of iron-doped PbTiO$_3$}
The theoretical studies of the system Fe$'$:PbTiO$_3$ have been
performed using an $2 \times 2 \times 2$ supercell, in which now
Ti is partly replaced by Fe. We have studied different
possibilities of ion arrangements for the defect structures. In a
model series A, we have replaced two Ti by Fe$'$, Fe$'_{\rm Ti}$,
and for charge compensation we have created one O vacancy, $V_{\rm
O}^{\bullet \bullet}$, that is the supercell has the composition
Pb$_8$Ti$_6$Fe$_2$O$_{23}$. In two other model series, we have
replaced only one Ti by Fe$'$. In the model series B, we consider
the neutral defect [Pb$_8$Ti$_7$FeO$_{23}$] and in the model
series C, the charged defect [Pb$_8$Ti$_7$FeO$_{23}]^+$, charge
compensation obtained by assuming a homogeneously charged
background.\par

The ion arrangement in the three model series A to C are as
follows. In model A1, the two Fe$'_{\rm Ti}$ are next nearest
neighbors and the $V_{\rm O}^{\bullet \bullet}$ is located between
these two Fe$'_{\rm Ti}$ forming a clustered defect oriented in
the $c$ direction. Model A2 differs from model A1 only by the
assumption that one of the Fe$'_{\rm Ti}$ is an isolated point
defect, whereas for model A3, we have assumed that all point
defects are isolated. Model A4 to A6 vary from model A1 to A3 by
the position of the oxygen vacancy, which is located in the $a$
(or $b$) direction (i.e., at O2 or O3 in figure
\ref{3D_structure}) rather than in the $c$ direction (O1 in figure
\ref{3D_structure}). For the systems of model series B
[Pb$_8$Ti$_7$FeO$_{23}$] and C [Pb$_8$Ti$_7$FeO$_{23}$]$^+$ with
only one Fe$_{\rm Ti}$ in the $2 \times 2 \times 2$ supercell we
have again studied the arrangement of $V_{\rm O}^{\bullet \bullet}
- {\rm Fe}'_{\rm Ti}$ neighbored in $a$ or $c$ direction (models
B4 or B1, B2 and C4 or C1, C2) as well as the arrangement with
both Fe$'_{\rm Ti}$ and $V_{\rm O}^{\bullet \bullet}$ as isolated
point defects (models B3 and C3). For the defect associate $V_{\rm
O}^{\bullet \bullet} - {\rm Fe}'_{\rm Ti}$ neighbored in the $c$
direction we distinguished between the case where the vacancy site
is at O1 (models B1 and C1) or at O1' (models B2 and C2) as the
distances of the O1 and O1' to Fe$'_{\rm Ti}$ are not the
same.\par

For all these models we have performed a structure optimization
varying the atomic positions but taking the lattice constants $a$
and $c$ fixed to the values obtained from the DFT results for the
pure host lattice given in table \ref{X_ray_results}. Considering
the total energy of the optimized structures it is concluded for
all three model series A to C that the oxygen vacancy in $c$
direction is always favored in energy, as compared to an
orientation along $a$. For models A1 to A6 the clustered defect
Fe$'_{\rm Ti} - V_{\rm O}^{\bullet \bullet} - {\rm Fe}'_{\rm Ti}$
along the $c$ direction (model A1) gives the lowest energy $E_{\rm
A1}$ followed by model A2 (Fe$'_{\rm Ti} - V_{\rm O}^{\bullet
\bullet}$ placed along the $c$ direction) with energy $E_{\rm
A2}$, where $E_{\rm A2} - E_{\rm A1} = 0.59$ eV per supercell.\par

Also for the model series with only one Fe$'_{\rm Ti}$ defect the
direction of the Fe$'_{\rm Ti} - V_{\rm O}^{\bullet \bullet}$
associate along $c$ (models B1 and C1) is preferred over the $a$
direction. More precisely, the most stable arrangement is the
arrangement were O1 is missing and not O1'. The models B1 and C1
are lower in energy than B2 and C2, namely $E_{\rm B2} - E_{\rm
B1} = 0.35$ eV and $E_{\rm C2} - E_{\rm C1} = 0.34$ eV. The
arrangements B4 and C4 with an orientation of the Fe$'_{\rm Ti} -
V_{\rm O}^{\bullet \bullet}$ dipole along the $a$ axis is less
stable than the favored arrangement within the model series by
$E_{\rm B4} - E_{\rm B1} = 0.45$ eV and $E_{\rm C4} - E_{\rm C1} =
0.38$ eV. The associated defect Fe$'_{\rm Ti} - V_{\rm O}^{\bullet
\bullet}$ is favourable compared to the isolated defects by
$E_{\rm B3} - E_{\rm B1} = 0.27$ eV and $E_{\rm C3} - E_{\rm C1} =
0.13$ eV. Applying the spin-polarised version of the LDA method
changes the energy by less than $0.02$ eV per $2 \times 2 \times
2$ supercell and therefore has not been pursued any further.\par

Next, the movement of the ions around the vacancy shall be
described. Qualitatively, we find the same trends for all models
considered here, namely the cation nearest to the vacancy moves
away from the vacancy site, which results in a shortening of the
metal oxygen bond opposite to the vacancy.\par

Comparing the undoped structure (figure \ref{structure} (a)) and
the defect structures (figure \ref{structure} (b)-(f)) one sees
that the position of the metal ion above the vacancy moves upwards
above the plane of the four oxygen ions which has already been
found for Ti$_{\rm Ti}$ in a previous calculation by Park and
Chadi \cite{Par98}. The displacement of the Fe ion from the
original Ti position in the undoped host lattice represented by
the distance $d$ is equal to $+46.6$ pm and $-11.6$ pm for the two
Fe$'_{\rm Ti}$ in model A1. For the structures with only one
Fe$'_{\rm Ti}$ we get the following values for $d$, $d_{\rm A2}$ =
$+32.9$ pm $d_{\rm B1} = +31.2$ pm, $d_{\rm C1} = +33.5$ and $d_{\ C2} = -6.8$ pm.\par

Considering the atomic arrangement and relaxation we finally point
out that even for the pure host the weak O - Ti bond of $243$ pm
compared to the other O - Ti bonds of $174$ and $198$ pm are not
described quite satisfactorily by the LDA results, see figure
\ref{structure}. This also holds for GGA (Generalized Gradient
Approximation) calculations which we have additionally performed
using the potential of Perdew and Wang \cite{Per92}. Therefore, we
have repeated all calculations under the constrain of leaving the
strain constant, $c/a=(c/a)_{exp}=1.0721$ by both LDA and GGA. From
these additional calculations we get the same main conclusions
described above, namely that the associate Fe$'_{\rm Ti} - V_{\rm
O}^{\bullet \bullet}$ in $c$ direction has the lowest energy and
compared to the original Ti and O positions the Fe$'_{\rm Ti}$ is
moving away from the vacancy V$_{\rm O}^{\bullet \bullet}$.\par

\begin{figure}[ht]
 \includegraphics[width=0.9\linewidth]{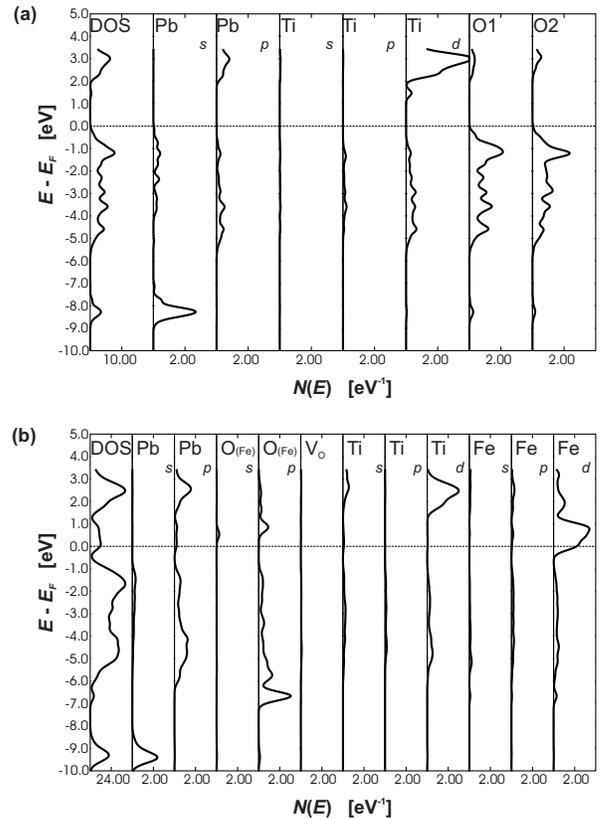}
 \begin{center}
  \caption{Density of states (DOS) for undoped PbTiO$_3$ (a) and Fe:PbTiO$_3$ (b). Total DOS and partial DOS of Pb, Ti($s,p,d$), O1-3($s,p$). The horizontal dotted line corresponds to the top of the valence band.}
  \label{band_structure}
 \end{center}
\end{figure}

The resulting electron states are presented in figure
\ref{band_structure}, in which the density of states (DOS) and the
partial DOS based on the Mulliken population analysis is plotted
for PbTiO$_3$ (figure \ref{band_structure} (a)). The lowest states
in figure \ref{band_structure} (a) are the 6$s$-like states of Pb.
Above these states we find the oxygen 2$p$-like states mixing with
the 6$s$- and 6$p$-like states of Pb and the 4$s$/4$p$-like states
of Ti. This indicates that the bonding has a distinct covalent
character as found also in previous calculations
\cite{Cohen,Par98,T11}. The lowest states of the conduction bands
are predominantly 3$d$-like states of Ti for pure PbTiO$_3$. For
Pb$_8$Ti$_6$Fe$_2$O$_{23}$ in figure \ref{band_structure} (b),
additional 1$s$ and 2$p$ basis functions centered at the vacancy
site have been used. No new states are found in the band gap,
which otherwise would indicate the existence of color centers at
the oxygen vacancy. Compared to figure \ref{band_structure} (a) we
get additional Fe like states and one can see that the highest
occupied states are now the 3$d$-like states of Fe. The oxygen,
for which the partial DOS are displayed in figure
\ref{band_structure}(b), is neighbored to Fe in the $c$-direction
and labelled as O(Fe$'$). Its 2$p_{\rm z}$-like states show a
common peak with the 3$d$-like states of Fe in an energy range of
$-7$ to $-6$ eV indicating an Fe - O bond.

\section{Discussion}
The most striking prediction from the {\it NSM} calculations is
that the Fe$^{3+}$ functional center in lead titanate forms a
charged Fe$'_{\rm Ti}-V_{\rm O}^{\bullet \bullet}$ defect
associate with an oxygen vacancy in the local oxygen octahedron,
the iron being substituted at the B-site of the perovskite ABO$_3$
lattice. The similarity of ionic radii between Ti$^{4+}$ at 68 pm
and Fe$^{3+}$ at 64 pm as compared to Pb$^{2+}$ with a radius of
124 pm supports this assignment. In fully or partly ionic
compounds, vacancies are charge balanced by other defects forming
an overall neutral system. It can be assumed that partial charge
compensation takes place at a nearest-neighbor oxygen site in the
octahedron, because the resulting Coulomb interaction is the most
important driving force for association. This assignment is in
accordance with first-principle calculations that predict the
binding of oxygen vacancies to acceptor-type impurities
\cite{Poe99}, as well as to our {\it DFT} investigations that
favor defect associates. Experimental findings indicate, however,
that for other acceptor centers in similar compounds, like
chromium-doped PbTiO$_3$ \cite{Erd03} and copper-modified
Pb[Zr$_{0.54}$Ti$_{0.46}$]O$_3$ \cite{RAE04ab}, no such
association is present. Alternatively, like in iron-doped
SrTiO$_3$, an equilibrium between 'free' Fe$'_{\rm Ti}$ centers
and Fe$'_{\rm Ti}-V_{\rm O}^{\bullet \bullet}$ defect associates
can be present \cite{Mer03}. With respect to the ionic mobility of
free oxygen vacancies, the iron - oxygen vacancy defect dipole
complex will be rather immobile in the ceramic. Hence, charge
transport will be considerably hindered. Furthermore, the charged
defect agglomerate may influence the poling properties by
providing pinning centers for domain walls \cite{Yan99}. Moreover,
dipolar defect complexes have been demonstrated to be able to also
pin the polarization of the surrounding crystal \cite{Poe99}. In
particular, oxygen-vacancy related defect dipoles have been shown
to be involved in voltage offsets leading to imprint failure
\cite{Pik95} and are suggested to play a crucial role in
electrical fatigue \cite{Poe99,Lupascu,Tag01}.\par

With respect to the reliability of the predictions obtained by the
semi-empirical {\it NSM} approach, the following points should be
mentioned. First, crystal distortions near the substituted ion and
contributions from ions or vacancies more distant than neighboring
ligands are neglected. Second, the main assumption of the {\it NSM} is
that the spin-Hamiltonian parameters result from individual
crystal field contributions of every nearest neighbor ion.
Finally, the model is based upon the calculation of single
electron-derived charge densities, whereas the FS interaction is
related to two-electron expectation values. Apparently, the {\it NSM}
has proven to yield reliable results for determining the second-
and fourth-rank finestructure parameters in $S$-state ions, probably
because the intrinsic parameters have been refined over several
decades using a large set of experimental data. If iron at a
B-site of the perovskite ABO$_3$ lattice can be considered as part
of this set, the structural data can be considered as
reliable.\par

Within these limits, the results can be interpreted as follows:
the Fe$^{3+}$ ion with the nearby oxygen vacancy probably has a
position near the center of the truncated oxygen octahedron, i.e.
at the (0.5000 0.5000 0.6288) position, whereas the second
position in accordance with {\it NSM} prediction can be discarded
because of {\it DFT}-based results.\par

As mentioned above, for all {\it DFT} calculations a $2 \times 2
\times 2$ unit cell was used. This size of the unit cell seems to
be too small to avoid interaction of the defects, and the
relaxation at the defect site could be smaller if larger unit
cells are used. However, for computation time reasons we could not
expand our unit cell. Nevertheless, it is most encouraging to find
that the local structure with respect to Fe$^{3+}$ positioning,
predicted by {\it DFT} calculation and by the semi-empirical model
are quite similar. The agreement is surprisingly good and might be
fortuitous, considering the rather small supercell accomplished by
an effective 12.5 \% iron doping \par

The quantum mechanical simulations show, that the association of
the defects is preferred over the arrangement as isolated point
defects. Additionally, the arrangements for an orientation of the
defect associate along the $c$-axis are more stable than the ones
along the $a$ and $b$-axes, confirming the conclusion drawn from
the axial symmetry of the FS tensor \cite{RAE04c}. The off-center
position of the Fe ion in the oxygen octahedron leads to two
Fe$'_{\rm Ti}-V_{\rm O}^{\bullet \bullet}$ distances in c
direction and therefore to two different positions of the oxygen
vacancy. According to the DFT investigations the vacancy at the O1
position in figure \ref{3D_structure} is energetically favored, in
which the long O1'- Fe$'_{\rm Ti}$ distance decreases
significantly.\par

Finally, because the observed EPR spectra \cite{RAE04c} can almost
exclusively be interpreted in terms of  Fe$'_{\rm Ti}-V_{\rm
O}^{\bullet \bullet}$ associates, one can conclude that no 'free'
Fe$^{3+}$ signals, i.e., iron ions without associated $V_{\rm
O}^{\bullet \bullet}$ and hence with considerably smaller FS
values, are present. Considering the condition for overall charge
compensation there is a charge mismatch, because the Fe$^{3+}$ ion
substituting for the Ti$^{4+}$ ions are singly negative charged
(Fe$'_{\rm Ti}$), whereas the associated oxygen vacancy is doubly
positive charged ($V_{\rm O}^{\bullet \bullet}$) with respect to
the neutral lattice. Hence, additional mechanisms for charge
compensation have to be discussed. Candidates for charge
compensation are either free electrons ($e'$) trapped in the
lattice, lead vacancies $(V''_{\rm Pb})$, and the formation of
positively charged cations, such as Pb$^{2+}$ $\rightarrow$
Pb$^+$, Ti$^{4+}$ $\rightarrow$ Ti$^{3+}$, or Fe$^{3+}$
$\rightarrow$ Fe$^{2+}$. However, there no evidence for color
centers in the {\it DFT} calculations and the variable valency
ions Pb$^+$, Ti$^{3+}$ are paramagnetic with $g$-values at about
$g \approx 2.0$ to 1.9, which were not observed in the EPR spectra
and thus can be excluded. Since intrinsic double negatively
charged lead vacancies $V''_{\rm Pb}$ have been suggested as
additional charge compensation \cite{War9697}, the overall
electro-neutrality condition for iron-modified lead titanate thus
is proposed to be given by

\begin{equation}
 [V_{\rm O}^{\bullet \bullet}] = [{\rm Fe}'_{\rm Ti}] + \frac{1}{2} [V''_{\rm Pb}]
\end{equation}

This model is supported by the inherent loss of PbO during
processing, for which reason a natural intrinsic $V_{\rm
O}^{\bullet \bullet}-V''_{\rm Pb}$ di-vacancy pair was proposed
\cite{Kee98}, its existence, however, being  currently
controversially discussed on the basis of {\it DFT} calculations
\cite{Poe00,Coc04}.\par

In principle, charge-compensation could also be obtained by the
creation of Fe$'_{\rm Ti}-V_{\rm O}^{\bullet \bullet}-$Fe$'_{\rm
Ti}$ defect associates without the need for lead vacancies at all.
The existence of such a defect structure is also supported by the
{\it DFT} results, predicting this arrangement as of lowest
energy. In this structure, the {\it DFT} calculations predicts two
different iron sites, for which reason two Fe$^{3+}$ EPR spectra
with different FS parameters are expected. Using the calculated
displacements of $d = +46.6$ pm and $-11.6$ pm, the corresponding
{\it NSM} estimates are $b_2^0 = 9.6$ GHz and $b_2^0 = -34.7$ GHz,
respectively. These values differ considerably from the
experimentally obtained one, and for this reason this model can be
discarded. Furthermore, for samples with low iron doping, this
prediction of an Fe$'_{\rm Ti}-V_{\rm O}^{\bullet
\bullet}-$Fe$'_{\rm Ti}$ defect associate might not be upheld.
Finally, from the present EPR experiment there is no further
evidence for this hypothesis, because no indication for strongly
dipolar coupled iron centers is detected in the EPR spectra.

\section{Conclusion}
In summary, a charged Fe$'_{\rm Ti}-V_{\rm O}^{\bullet \bullet}$
defect associate in lead titanate has been identified and its
microscopic structure has been determined based on a comparison of
FS data with results of semi-empirical {\it NSM} and {\it DFT}
calculations. The refined structure comprises information about
the structural relaxation around the iron functional center in
lead titanate. In the model presented, Fe$^{3+}$ is substituted as
an acceptor center at the perovskite B-site with a directly
coordinated oxygen vacancy. The position of the iron ion in the
ferroelectric phase is found to be almost centered in the unit
cell, different from the bulk B-site Ti$^{4+}$ ions in the
ferroelectric phase that are considerably displaced, but similar
to the B-site positions in the paraelectric phase.\par

The orientation of the Fe$'_{\rm Ti}-V_{\rm O}^{\bullet \bullet}$
defect dipole is found to be along the crystallographic $c$-axis.
This can be concluded from EPR results which indicate axial
symmetry at the iron center. Any other orientation of the defect
dipole would result in an FS tensor of lower than axial symmetry.
Furthermore, the {\it DFT} calculations confirm this assignment by
showing that the total energy of the arrangement along the
$c$-axis is more stable than the ones along the $a$ and $b$-axes.
Using high-resolution synchrotron powder diffraction, a quite
large $c/a$-ratio of 1.0721 was measured, again pointing to the
uniqueness of structural relaxation along \emph{c}.\par

The impact of iron-doping in lead titanate on the macroscopic
piezoelectric properties can now be rationalized from a
microscopic point of view, because the iron - oxygen vacancy
defect dipole complex will be rather immobile in the ceramic as
compared to the ionic mobility of free oxygen vacancies. Thus,
charge transport will be considerably hindered and the charged
defect agglomerate may furthermore influence domain-wall motion
and poling properties by providing pinning domain walls.

\section{Acknowledgments}
This investigation has been financially supported by the DFG
priority program 1051 {\it 'High-Field EPR in Biology, Chemistry
and Physics'} and center of excellence 595 {\it 'Electrical
Fatigue in Functional Materials'}. The authors are grateful to
helpful comments of the referees made during the review process.
Furthermore, we thank Dr. D.J. Keeble for sending us a preprint
about his X-band EPR study of Fe$^{3+}$ centers in lead titanate
single crystals prior to publication.

\end{document}